\documentclass[runningheads]{llncs}

\usepackage{cite}
\usepackage{amsmath,amssymb,amsfonts}
\usepackage{algorithmic}
\usepackage{graphicx}
\usepackage{textcomp}
\usepackage{xcolor}
\def\BibTeX{{\rm B\kern-.05em{\sc i\kern-.025em b}\kern-.08em
    T\kern-.1667em\lower.7ex\hbox{E}\kern-.125emX}}
    
\usepackage{tikz}    
\usepackage{url}
\usepackage{caption}
\usepackage{subcaption}

\usepackage{hyperref}

\usepackage{array,booktabs,calc}
\usepackage{dsfont}
\usepackage{eqnarray,amsmath}
\usepackage[ruled,lined]{algorithm2e}
\usepackage{bbold}
\usepackage[titles]{tocloft}
\usepackage{setspace}

\newcommand{\IN}{\!\in\!}

\newcommand{\mcCTMC}[1][\phantom{}]{\mathcal{M}_{#1}}
\newcommand{\mcpCTMC}{(\mcCTMC[\theta])_{\theta\in \Theta}}
\newcommand{\eltS}{\mathbf{s}}

\newcommand{\veccyln}[2][{[n,m]}]{C({#2}, \eltS_{0:{#2}}, {#1}^{#2})}

\newcommand{\meantp}{\bar{t_p}}
\newcommand{\vartp}{s^2_{t_p}}
\newcommand{\obsmeantp}{\meantp^{(obs)}}

\newcommand{\aut}{\mathcal{A}}
\newcommand{\autper}{\mathcal{A}_{per}}
\newcommand{\autpertpobs}{\mathcal{A}^{\bar{t_p}^{(obs)}}_{per}}

\newcommand{\Unif}[2]{\mathcal{U}(#1,#2)}

\newcommand{\eps}{\epsilon}

\newcommand{\piabceps}[1][\phantom{}]{\pi_{ABC}^{\eps_{#1}}}

\usetikzlibrary {plotmarks,calc,arrows,shapes,snakes,automata,backgrounds,petri,positioning,fit}
\usetikzlibrary{automata,shapes.multipart} 
\usetikzlibrary{arrows,petri}
\usetikzlibrary{arrows}

\definecolor{ForestGreen}{RGB}{34,139,34}

\definecolor{ForestGreen}{RGB}{34,139,34}

\usepackage{textcomp}
\usepackage{listings}

\definecolor{prismident}{HTML}{cc0000}
\definecolor{prismnum}{HTML}{0000cc}
\definecolor{prismcomment}{HTML}{228B22}
\definecolor{prismpreproc}{HTML}{a020f0}

\lstdefinestyle{myFramedSmallCPP} {
    language=[11]C++,
    basicstyle=\small\color{black},
    keywordstyle=\bfseries,
    commentstyle=\color{ForestGreen},
    showstringspaces=false,
    escapechar=\#,
    numbers=none,
    frame=single,
    morekeywords = {final, override},
    literate={~} {$\sim$}{1}
}

\lstdefinestyle{myFramedFootNotePRISMnoNum} {
    morekeywords=[2]{1, 2, 3, 4, 5, 6, 7, 8, 9, 0},
    keywordstyle=[2]{\color{blue}},
    basicstyle=\footnotesize,
    upquote=true,
    keywordstyle=\bfseries,
    identifierstyle=\color{red},
    numberstyle=\color{blue},
    commentstyle=\color{ForestGreen},
    showstringspaces=false,
    escapechar=\#,
    numbers=none,
    frame=single,
    morekeywords = {module, formula, init, endmodule, reward, endreward, dtmc, ctmc, mdp, ', max, min},
    literate={~} {$\sim$}{1}
}

\lstdefinestyle{myFramedTinyPRISMnoNum} {
    morekeywords=[2]{1, 2, 3, 4, 5, 6, 7, 8, 9, 0},
    keywordstyle=[2]{\color{blue}},
    basicstyle=\tiny,
    upquote=true,
    keywordstyle=\bfseries,
    identifierstyle=\color{red},
    numberstyle=\color{blue},
    commentstyle=\color{ForestGreen},
    showstringspaces=false,
    escapechar=\#,
    numbers=none,
    frame=single,
    morekeywords = {module, formula, init, endmodule, reward, endreward, dtmc, ctmc, mdp, ', max, min},
    literate={~} {$\sim$}{1}
}

\lstdefinestyle{myFramedFootNotePRISM} {
    morekeywords=[2]{1, 2, 3, 4, 5, 6, 7, 8, 9, 0},
    keywordstyle=[2]{\color{blue}},
    basicstyle=\footnotesize,
    upquote=true,
    keywordstyle=\bfseries,
    identifierstyle=\color{red},
    numberstyle=\color{blue},
    commentstyle=\color{ForestGreen},
    showstringspaces=false,
    escapechar=\#,
    numbers=left,
    frame=single,
    morekeywords = {module, formula, init, endmodule, reward, endreward, dtmc, ctmc, mdp, ', max, min},
    literate={~} {$\sim$}{1}
}

\lstdefinestyle{myFramedTinyPRISM} {
    morekeywords=[2]{1, 2, 3, 4, 5, 6, 7, 8, 9, 0},
    keywordstyle=[2]{\color{blue}},
    basicstyle=\tiny,
    upquote=true,
    keywordstyle=\bfseries,
    identifierstyle=\color{red},
    numberstyle=\color{blue},
    commentstyle=\color{ForestGreen},
    showstringspaces=false,
    escapechar=\#,
    numbers=left,
    frame=single,
    morekeywords = {module, formula, init, endmodule, reward, endreward, dtmc, ctmc, mdp, ', max, min},
    literate={~} {$\sim$}{1}
}

\begin{document}


\title{A Formal Approach For   Tuning Stochastic Oscillators}




\titlerunning{Tuning stochastic oscillators}

\author{
Paolo Ballarini\inst{1}
\and
Mahmoud Bentriou\inst{1}
\and
Paul-Henry Cournède\inst{1}}
\authorrunning{P. Ballarini et al.}
%
\institute{MICS, CentraleSup\'{e}lec, 
Universit\'{e} Paris-Saclay,  France. \email{paolo.ballarini@centralesupelec.fr}\\
\email{mahmoud.bentriou@centralesupelec.fr}\\
\email{paul-henry.cournede@centralesupelec.fr}
}

\maketitle

\begin{abstract}
Periodic recurrence is a  prominent  behavioural  of many  biological phenomena, including cell cycle and circadian rhythms. Although deterministic models are commonly used to represent the dynamics  of periodic phenomena, it is known that they are little appropriate in the case of systems in which \emph{stochastic noise} induced  by small population numbers is actually responsible for  periodicity. 
Within the stochastic modelling settings automata-based model checking approaches have  proven an effective means for  the analysis of oscillatory dynamics, the main idea being that of coupling a \emph{period detector} automaton with a  continuous-time Markov chain model of an alleged oscillator. 
In this paper we address a complementary aspect, i.e. that of assessing  the dependency  of oscillation related measure  (period and amplitude)  against the parameters of a stochastic oscillator. To this aim we introduce a framework  which, by  combining an Approximate Bayesian Computation scheme with a hybrid automata capable of quantifying how   \emph{distant} an instance of a stochastic oscillator is from matching a desired (average) period,  leads us to identify regions of the parameter space in which oscillation with given period  are highly likely. 
The method is demonstrated through a couple of case studies, including a model of the popular Repressilator circuit. 

\end{abstract}



\section{Introduction}
Oscillations are prominent dynamics at the core of many fundamental biological processes. They occur at different level and concern different time scales, ranging 
 from ion channels regulated transmission of intercellular electrical signal driving the heartbeat (with period in the order of one second), to  intracellular calcium oscillations triggering glycogen-to-glucose release in liver cells (with periods ranging from few seconds to few minutes~\cite{doi:10.1073/pnas.0506135103}), to gene-expression regulated  circadian cycle (with typical period of roughly 24 hours). 
 
 Mathematical modelling and computational methods have proved fundamental to gain a better   understanding of the complexity of the  mechanisms that regulate oscillations~\cite{Goldbeter2002ComputationalAT}. Although continuous-deterministic  models (i.e. ODEs)  are more often considered in the literature of biological oscillators, with systems characterised by low population numbers  (e.g.,  genetic circuits responsible for \emph{circadian rhythms}),  discrete-stochastic models are more appropriate~\cite{MacLaurin22}: when few molecules are involved the stochasticity of the system becomes important resulting in noisy periodic behavior. 
 Understanding the effect that an oscillator's parameters have on the quality of the oscillations is a relevant research problem.





\smallskip
\noindent
{\bf Contribution}. 
We introduce a methodology for calibrating stochastic models that exhibit a noisy periodic dynamics. 
Given a parametric stochastic oscillator   the methodology allows to identify regions of the parameter space with a positive probability of matching a  given target oscillation period. 
This entails integrating formal means for  noisy periodicity analysis within  a parameter inference approach, more specifically,      integrating a  \emph{period distance meter} (formally encoded as a Hybrid Automaton) within an Approximate Bayesian Computation scheme. Given some  period-related requirements (e.g. mean value and variance of the oscillation period) the trajectories sampled from the oscillator are classed through the distance period meter   leading us to the approximation of the corresponding posterior distribution, that is, to the identification of the region of the parameter space that are more likely to comply with the  considered period-related requirements.

\smallskip
\noindent
{\bf Paper organisation}. The paper is organised as follows: 
in Section~\ref{sec:background} we overview the background material our approach relies upon, that includes an overview of  HASL model checking and of ABC algorithms. 
In Section~\ref{sec:methods} we introduce our approach for calibrating stochastic oscillators w.r.t. to the oscillation period. The method is then demonstrated through experiments presented in Section~\ref{sec:experiments}. Conclusive remarks and future perspective are discussed in Section~\ref{sec:conclusion}.


\subsection{Related work} 
\smallskip
\noindent
{\bf Temporal logic based analysis  of stochastic oscillators}. 
The analysis  of oscillatory behaviour entails two complementary aspects: determining whether a model exhibits recurrent patterns (detection) and assessing relevant periodicity indicators   (e.g. period and amplitude of oscillations). In the continuous-deterministic   settings these can be achieved through a combination of mathematical approaches including structural analysis of the corresponding ODE system and stability analysis  of its \emph{steady-state} solutions~\cite{Goldbeter2002ComputationalAT}. As those approaches clearly do not apply in the discrete-stochastic settings researchers progressively started looking at alternatives such as the adaptation of  model checking to the analysis of periodicity~\cite{DBLP:conf/rocks/AndreychenkoKS12}. 
Detection of sustained oscillations through model checking requires identifying temporal logical   formulae that  single out  infinite  cyclic (non-constant) behaviour. 
Seminal ideas introduced for (non-probabilistic) \emph{transition systems} models~\cite{ChabrierRivier2004ModelingAQ} and  further developed in~\cite{Ballarini2009AnalysingChecking}  included Computational Tree Logic (CTL)~\cite{DBLP:conf/popl/ClarkeES83} \emph{qualitative} specifications such as 
\begin{eqnarray}
\label{ctloscillexist}
EG ( ((X_i=k) \Rightarrow EF(X_i\neq k)) \land ((X_i\neq k) \Rightarrow EF(X_i= k)))\\
\label{ctloscillall}
AG ( ((X_i=k) \Rightarrow EF(X_i\neq k)) \land ((X_i\neq k) \Rightarrow EF(X_i= k)))
\end{eqnarray}
which  encode \emph{infinite alternation} by demanding that for  ``at least one path''  (\ref{ctloscillexist}) or for ``all paths''  (\ref{ctloscillall}) of a given model a state condition  $X_i=k$\footnote{the population of species $i$ is $k\in\mathbb{N}$} can be met  and then left (and, inversely, whenever it is not met  $X_i\neq k$  it is possible to eventually meet it) infinitely often. 
In the stochastic,  continuous time Markov chains (CTMCs), settings the Continuous Stochastic Logic (CSL)~\cite{Aziz1996CSL} counterpart of  (\ref{ctloscillall}) leads to the following qualitative probabilistic   formulae (introduced in~\cite{Ballarini2009AnalysingChecking}):
\begin{equation}
\label{csloscill}
P_{=1} ( ((X_i=k) \Rightarrow P_{>0}(X_i\neq k)) \land ((X_i\neq k) \Rightarrow P_{>0}(X_i= k)))
\end{equation}
which identifies those states of a CTMC for which the probability of outgoing paths that infinitely alternate between  $X_i=k$ and $X_i\neq k$ states adds up to 1. If formulae such as  (\ref{csloscill}) can be used to rule out CTMCs that do not oscillate sustainably (as they contain at least an absorbing state), 
they fall short w.r.t. effectively detecting sustained oscillators as 
 they are satisfied by any \emph{ergodic} CTMC regardless whether it actually exhibits sustained oscillations. 
The problem is that,  in the stochastic settings, a logical characterisation such as~(\ref{csloscill}) is too weak to allow for distinguishing between models  that gather probability mass on  actual oscillatory  paths (of given amplitude and period) from those  whose probability mass is concentrated on infinite  noisy fluctuations which do not  correspond with actual oscillations. Further developments included employing the probabilistic version of the linear time logic (LTL)~\cite{DBLP:conf/focs/Pnueli77} to characterise oscillations based on a notion of \emph{noisy monotonicity}~\cite{Ballarini20102019}: although an improvement w.r.t. to the limitation of the CSL based characterisation of periodicity this approach  still fails to satisfactorily  treat the oscillation detection problem. 



\smallskip
\noindent
{\bf Automata-based analysis  of stochastic oscillators}. 
To work around the limited suitability of temporal logic approaches  Spieler~\cite{Spieler13} proposed to employ a single-clock deterministic timed automata (DTA) as \emph{noisy period detector}. 
The idea is that, through synchronisation with a CTMC model, such DTA is used to accept \emph{noisy periodic} paths described as those that infinitely alternate between crossing a lower  threshold $L$, and a higher threshold $H$ (hence corresponding to  fluctuations of minimal amplitude $H\!-\!L$) and by imposing that such fluctuations should happen with a period falling in a chosen interval $[t_p^{min},t_p^{max}]$ (with $H$, $L$,  $t_p^{min}$ and $t_p^{max}$ being parameters of the DTA). The issues that CSL detector formulae such as~(\ref{csloscill}) suffer from are overcome, as by properly settings thresholds $L$ and $H$ the DTA rules out non-oscillating ergodic CTMC models. 
The detection procedure boils down to computing the probability measure of all CTMC paths that are accepted by the DTA which is achieved through numerical procedures (requiring the construction of the CTMC x DTA product process).
Spieler's original idea has then  further evolved by resorting to the more expressive hybrid automata as detectors of periodicity~\cite{DBLP:journals/tcs/BallariniD15,ballsttt15}. This allowed,  on one hand, to account for more sophisticated oscillation related indicators such as the variance (other than the mean value) of the period of a stochastic oscillator and also to develop an alternative, \emph{peaks detector}, automaton which, differently from Spieler's \emph{period detector}, does not depend on the chosen $L$ and $H$ thresholds. 



\smallskip
\noindent
{\bf Parametric verification of stochastic models}. 
Parametric verification of a probabilistic model 
is concerned with combining parameter estimation techniques with stochastic model checking, i.e. with studying how the stochastic model checking problem for a property $\Phi$ is affected by the parameters $\theta$ upon which a probabilistic model ${\cal M}_\theta$ depends. 
A number of approaches have been proposed in the literature such as~\cite{Brim2013ExploringParameterSpace}, in which a bounded approximation of parameter space fulfilling a CSL~\cite{Aziz1996CSL} threshold formula is efficiently determined through an adaptation of  \emph{uniformisation}, or also~\cite{Han2008ParameterSynthesis,Ceska2014ParameterSynthesis} and, more recently,  a novel ABC-based method~\cite{molyneux2020bayesian} that is based on observations even solves parameter inference and statistical parameter synthesis in one go. 

The framework we introduce in this paper, on the other hand, tackles the \emph{estimation problem} in stochastic model checking and is in line with the works of Bortolussi \emph{et al.}~\cite{BortolussiMS16}, where the so-called smoothed model checking (Smoothed MC) method to estimate the satisfaction probability function of parametric Markov population models is detailed.
The goal here is to obtain a good estimate of the so-called satisfaction probability function of the considered property $\Phi$ w.r.t. to model's parameter space $\Theta$. In our case we rely on hybrid automata-based adaptations of Approximate Bayesian Computation (ABC) schemes~\cite{bentriou2019,DBLP:journals/tcs/BentriouBC21} to estimate the satisfaction probability function, similarly to~\cite{DBLP:conf/cmsb/MolyneuxA20} where  ABC algorithms are combined with statistical model checking to approximate the satisfaction function of non-nested CSL reachability formulae. 

\section{Background}
\label{sec:background}
We briefly overview 
the notion of chemical reaction network (CRN) models, of  Hybrid Automata Stochastic Logic (HASL) model checking and the Approximate Bayesian Computation (ABC) scheme for parameter estimation that Section~\ref{sec:methods} relies on. 

\smallskip
\noindent
{\bf Chemical Reaction Network and discrete-stochastic interpretation}. We consider models that describe the time evolution of  biochemical species $X_1,X_2,\ldots$ which interact through a number of reactions $R_1,R_2,\ldots$ forming a so-called  chemical reaction network (CRN) expressed as a system of chemical equations with the following form:
\begin{equation*}
R_j:\sum_{i=1}^n a^-_{ij} X_i \xrightarrow{k_j} \sum_{i=1}^n a^+_{ij} X_i
\end{equation*}
where $a^-_{ij}$ ($a^+_{ij}$), are the stoichiometric coefficients of the \emph{reactant} (\emph{product}), species and $k_j$ is the kinetic constant of the $j$-th reaction $R_j$ of the CRN. 
We assume the discrete stochastic interpretation of CRNs models, i.e.  species  $X_i$ represent  molecules counting (rather than concentrations) hence models consist of  countable states  $x=(x_1,\ldots x_n)\in\mathbb{N}^n$ representing the combined populations count while the state transitions consist of time-delayed jumps governed by a probability distribution function. 
In case  of Exponentially distributed jumps the underlying class of models is that of continuous-time Markov chains (CTMCs), conversely in case of generically distributed jumps we refer to underlying class of models as  of  discrete event stochastic process (DESP).
A path of a DESP model is a (possibly infinite) sequence of  jumps denoted $\sigma\!=\! x^0\xrightarrow[R^0]{t_0} x^1\xrightarrow[R^1]{t_1} x^2\xrightarrow[R^2]{t_2}  \ldots $
with $x^i$ being the $i$-th state, $t_i\!\in\!\mathbb{R}_{>0}$ being the sojourn-time in the $i$-th state  and $R^i$ being the reaction whose occurrence issued  the ($i\!+\!1)$-th jump. Notice that paths of a DESP are \emph{c\`adl\`ag} (i.e. step) functions of time (e.g.  Figure~\ref{fig:LHA_toy1}). A DESP  depend on a  $p$-dimensional vector of parameters $\theta=[\theta_1,\ldots,\theta_p]\in\Theta\subseteq\mathbb{R}^p$. which, in the context of this paper, affect the kinetic rate
of the reaction channels. For $\theta\in\Theta\subseteq\mathbb{R}^p$ we denote ${\cal M}_\theta$ the corresponding DESP model.  
Given  ${\cal M}_\theta$   we let $Path({\cal M}_\theta)$ denote the set of paths of ${\cal M}_\theta$.   It is well known that a DESP model ${\cal M}_\theta$  induces a probability space over the set of events $2^{Path({\cal M}_\theta)}$, where the probability of a set of paths $E\in 2^{Path({\cal M}_\theta)}$ is given by the probability of their common finite prefix~\cite{Baier2003}. 
For ${\cal M}_\theta$  an $n$-dimensional DESP  population model, we denote  $dom({\cal M}_\theta)$  its state space, and  $dom_i({\cal M}_\theta)$  
the projection of $dom({\cal M}_\theta)$ along the $i^{th}$ dimension $1\!\leq\! i\leq n$ of ${\cal M}_\theta$. For $\sigma$  a path of an  $n$-dimensional ${\cal M}_\theta$ we denote $\sigma_i$ its projection along the $i$-th dimension.

\begin{figure}
    \centering
    \includegraphics[scale=0.25]{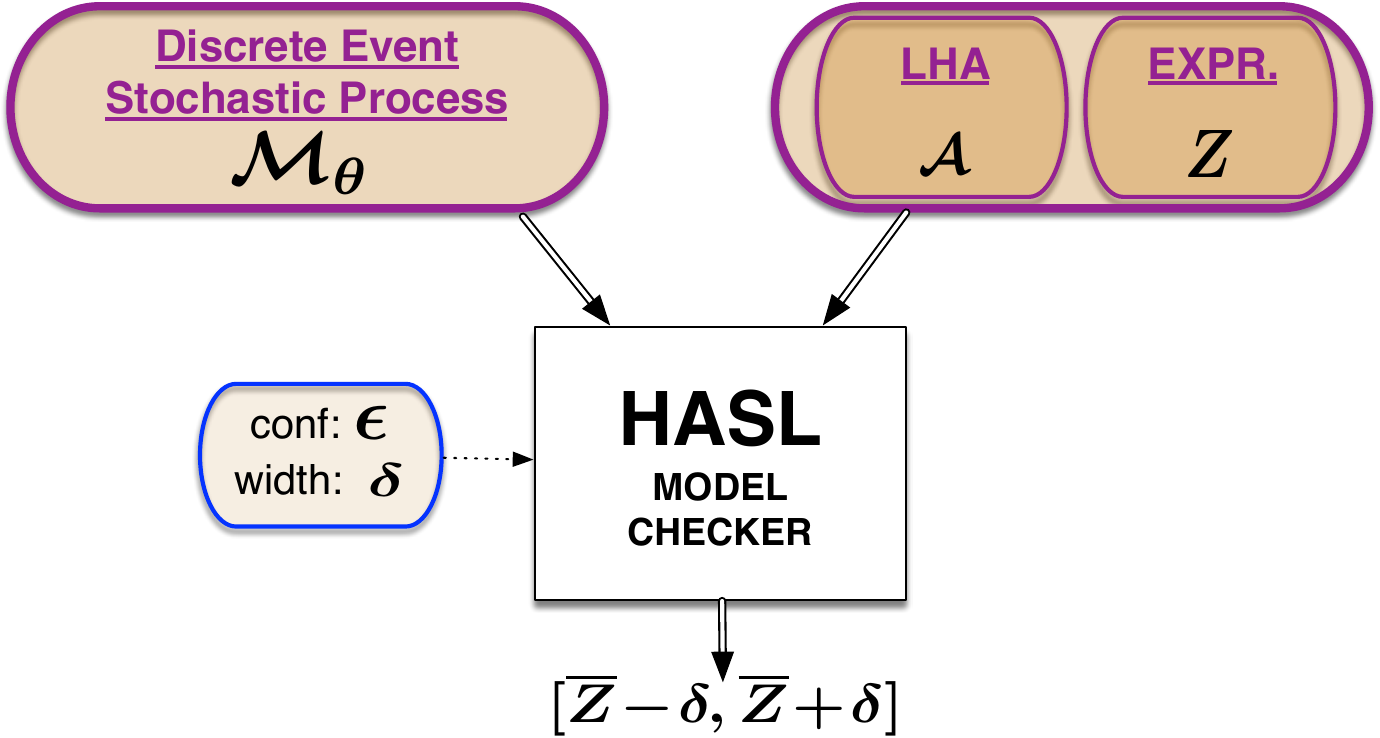}
    \caption{HASL-SMC schema: sampled paths are filtered by a LHA and the accepted ones used for a confidence interval estimate of the target measure.}
    \label{fig:hasl}
\end{figure}

\subsection{HASL model checking}
\label{sec:hasl}
The procedure for tuning stochastic oscillators we introduce in the remainder relies on the HASL statistical model checking (SMC) framework. We quickly overview the basic principles of HASL-SMC referring the reader to the literature~\cite{BALLARINI201553} for more details.
HASL-SMC is a framework  for assessing properties of stochastic model expressed by means of a linear hybrid automaton (LHA). It consists of an iterative procedure which 
 takes 3 inputs (Figure~\ref{fig:hasl}): a  parametric DESP model ${\cal M}_\theta$, a LHA ${\cal A}$,  and a target expression denoted $Z$ and outputs a confidence interval estimation of the mean value $\hat{Z}$ of the target measure. 
  At each iteration the procedure samples a path $\sigma\in Path({\cal M}_\theta\times\aut)$ of the product process ${\cal M}_\theta\times\aut$ (whose formal semantics is given in~\cite{BALLARINI201553}), that means that a path $\sigma\in Path({\cal M}_\theta)$ is sampled and synchronised \emph{on-the-fly} with $\aut$ leading to either acceptance or rejection of $\sigma$. In the synchronisation process,  relevant statistics of the simulated path are computed  and are used to define the target measure $Z$ of interest (see~\cite{BentriouPhD}). 
 This let HASL-SMC be  a very expressive formalism that allows for sophisticated signal-processing-like  analysis of a model. 
 
 
 \medskip

\smallskip
\noindent
{\bf Synchronisation of a model with a LHA}. 
The semantics of the synchronised process ${\cal M}\times{\cal A}$ naturally yields a stochastic simulation procedure which is implemented by the HASL model checker.
For the sake of space here we only provide an intuitive (informal) description of such  synchronisation  based on the example given in Figure~\ref{fig:LHA_toy1}. 
 An LHA consists of a set of locations $L$ (with at least one initial and one final location) a set of real-valued variables $V$ 
whose value may evolve, according to a flow function which map  each location to a (real-valued) rate of change to the variables, during the sojourn in a location. 
Locations change  through transitions denoted $l\xrightarrow{\gamma,E',U}l'$ where $\gamma$ is an enabling guard (an inequality built on top of variables in $V$), $E'$ is either set of events names (i.e.  the transition is \emph{synchronously} traversed  on occurrence of any  reaction in $E'$ occurring in the  path being sampled) or $\sharp$ (i.e.  the transition is \emph{autonomously} traversed without synchronisation) and $U$ are the variables' update. 
A state of ${\cal M}_\theta\times\aut$ is a 3-tuple $(s,l,\nu)$, with $s$ the current state of ${\cal M}_\theta$, $l$  the current location of $\aut$ and $\nu\in\mathbb{R}^{|V|}$ the current values of $\aut$'s variables. Therefore  if 
$\sigma: s\xrightarrow[e_1]{t_1}s_1\xrightarrow[e_2]{t_2}s_2\ldots $
is a path of ${\cal M}_\theta$ the corresponding path  $\sigma\times{\cal A}\in Path({\cal M}_\theta\times\aut)$  may be
$\sigma\times{\cal A}: (s,l,\nu)\xrightarrow[e_1]{t_1}(s_1,l_1,\nu_1)\xrightarrow[\sharp]{t^*_1}(s_1,l_2,\nu_2)\xrightarrow[e_2]{t_2}(s_2,l_3,\nu_3)\ldots $
where, the sequence of transitions $e_1$ and $e_2$ observed on $\sigma$ is interleaved with an autonomous transition (denoted $\sharp$) in the product process: i.e. from state $(s_1,l_1,\nu_1)$ the product process jumps to state $(s_1,l_2,\nu_2)$ (notice that state of ${\cal M}_\theta$ does not change) before continuing mimicking $\sigma$.

\begin{figure*}[h]
\begin{center}
\begin{tabular}{|c|c|}
\hline
 ${\cal M}$ &  ${\cal A}$ \\
 \hline
\begin{tabular}{l}
$R_1: A + B \xrightarrow{r_A} 2 B
$ \\
$R_2: B + C \xrightarrow{r_B} 2 C
$ \\
$R_3: C + A \xrightarrow{r_C} 2 A
$ \\
\end{tabular}
&
\begin{tabular}{l}
\scalebox{0.9}{
\begin{tikzpicture}[scale=.8,minimum width=1cm]
\renewcommand{\arraystretch}{.7}
   \everymath{\scriptstyle}
   
\draw (-3,1.5) node[initial,left, initial text=,draw,rounded corners] (l0) { \begin{tabular}{@{}c@{}}\underline{$l_0$}\\  $\dot{t}:1$ \\ $\dot{x_1}:A$\end{tabular} };
\draw (0.25,1.5) node[draw,rounded corners,accepting] (l1) { \begin{tabular}{@{}c@{}}\underline{$l_1$}\end{tabular} }; 

\draw [-latex'] (l0) -- (l1) node [midway, above,sloped] {$\sharp$,$t=T$,$\{x_1/=4\}$ };
\draw [-latex'] (l0) .. controls +(115:15mm) and +(75:15mm) .. (l0) node [midway ,above] { $\{R_1\}$,$t<4$,$\{n_2++\}$ };
\draw [-latex'] (l0) .. controls +(-115:15mm) and +(-75:15mm) .. (l0) node [midway ,below] { $ALL\setminus\{R_1\}$,$t< 4$,$\emptyset$ };




{}; 
 
\end{tikzpicture}}
\end{tabular}   \\ 
\hline \hline
\multicolumn{2}{|c|}{
\begin{tabular}{c}
{\centering synchronisation of ${\cal M}$ and ${\cal A}$}
\\
\includegraphics[scale=0.3]{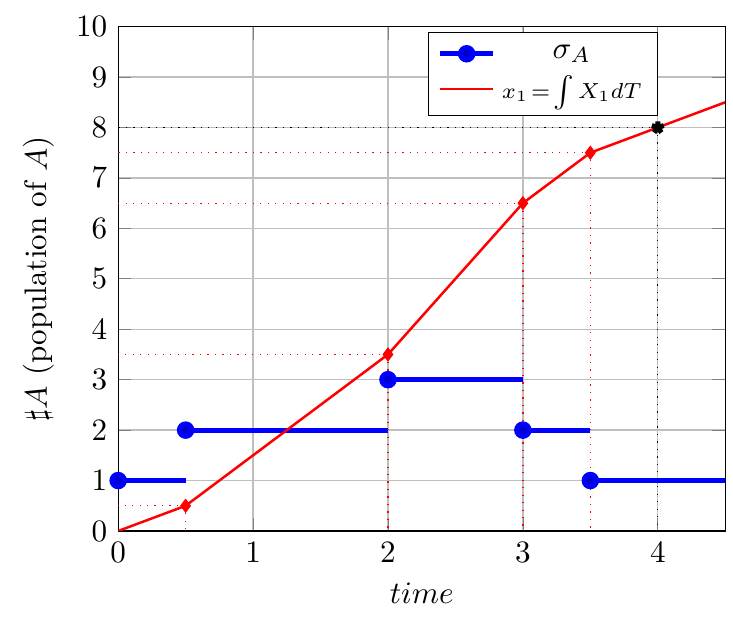}
\\
\end{tabular}} \\
\multicolumn{2}{|l|}{
\begin{tabular}{l}
{\scriptsize \underline{\bf path of ${\cal M}$.}}
\begin{scriptsize}
$\sigma:(1,2,3)\xrightarrow[R_3]{0.5}(2,2,2)\xrightarrow[R_3]{1.5}(3,2,1)\xrightarrow[R_1]{1}(2,3,1)\xrightarrow[R_1]{0.5}(1,4,1)
$
\end{scriptsize}\\ 
{\scriptsize 
\underline{{\bf corresponding path of ${\cal M}\times{\cal A}$.}}}\\
{\scriptsize $\sigma\times{\cal A}:((1,2,3),l_0,[0,0,0])\xrightarrow[R_3]{0.5} ((2,2,2),l_0,[0.5,0.5,0])\xrightarrow[R_3]{1.5} ((3,2,1),l_0,[2,3.5,0]) $}
\\ 
{\scriptsize \hskip 4ex
$\xrightarrow[R_1]{1}((2,3,1),l_0,[3,6.5,1])
  \xrightarrow[R_1]{0.5} ((1,4,1),l_0,[3.5,7.5,2])\xrightarrow[\sharp]{0.5} ((1,4,1),l_1,[4,8/4,2])$}
\\
\end{tabular}
}\\
\hline
\end{tabular}  

\caption{Example of synchronisation of a CRN model with a LHA.}
\label{fig:LHA_toy1}
\end{center}
\end{figure*}
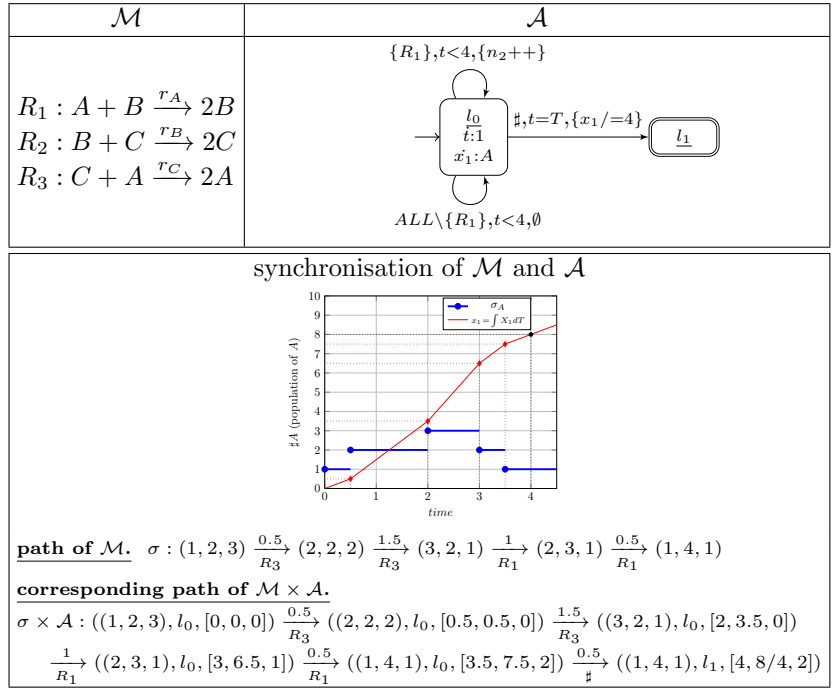
\begin{example}
\label{ex1}
Figure~\ref{fig:LHA_toy1} depicts an example of synchronisation between a path of a  toy  3-species, 3-reactions CRN model (left),  with a  2-locations LHA (right) designed for assessing properties of it. 
The LHA  locations are $l_0$ (initial)  and $l_1$ (final) while its  variables are $V=\{t,x_1,n_2\}$ with $t$ a clock variable,  $x_1$ a real valued variable (for measuring the average population of $A$) and  $n_2$  an integer variable (for counting the number of occurrences of the $R_1$ reaction). 
 While in $l_0$ the  value of variables  evolves according to their flow,  which is constant and equal to 1 for clock $t$, while is given by the current population of $A$ for $x_1$ (therefore   $x_1$ measures the integral of  population  $A$ along the synchronising path). 
The synchronisation of a path $\sigma$ with $\aut$ works  as follows:  $\aut$  stays in the initial location $l_0$ up until at $t=4$ the autonomous transition  the synchronisation with $\sigma$ ends as soon as the \emph{autonomous} transition $l_0\xrightarrow{\sharp,t=4,\{x_1/=4\}}l_1$  becomes enabled (guard $t=4$ gets true) hence is fired (by definition autonomous transitions have priority over \emph{synchronised} transitions).  As long as $t<4$ the LHA is in  $l_0$ where it synchronises with the occurrences of the reactions of the CRN model: on occurrence of  $R_1$ the ${\cal M}\times{\cal A}$ transition $l_0\xrightarrow{\{R_1\},t<4,\{n_2++\}}l_0$ (\emph{synchronised} on event set $\{ R_1\}$) is fired hence increasing the counter $n_2$, whereas on occurrence of any other reaction   transition  $l_0\xrightarrow{ALL\setminus\{R_1\},t<4,\emptyset}l_0$ (\emph{synchronised} on event set $ALL\setminus\{R_1\}$, where $ALL$ denotes all reactions of the CRN)  fires without updating any variable. Finally on ending the synchronisation with $\sigma$ variable $x_1$ is update to $x_1/4$ which  corresponds to average population of  $A$ observed over the time interval $[0,4]$. Such a LHA can therefore be used  (through iterated synchronisation with a sufficiently large number of trajectories) for estimating the confidence interval of random variables such as the ``average population of $A$'' as well as the  ``number of $R_1$ occurrences'' observed over  time interval $[0,4]$. 
An example of synchronisation between a path $\sigma$   consisting of 2 occurrences of $R_3$ (at $t=0.5$, $t=2.0$ respectively) followed by 2 occurrences of $R_1$ (at $t=3.0$ and $t=3.5$ respectively) and for which  $(A_0,B_0,C_0)=(1,2,3)$ is assumed as the initial state is depicted at bottom of Figure~\ref{fig:LHA_toy1}. The  synchronised path shows the combined evolution of the model's state, the LHA location and the value of LHA variables.
Notice that when synchronisation ends ($l_1$ is reached) variable $x_1$ is assigned with $8/4$ which  indeed is the mean population of $A$ along $\sigma$ until $t=4$. 
\end{example}

\subsection{Approximate Bayesian Computation}
\label{sec:abc}
The framework for tuning of stochastic oscillators we present in the remainder relies on the integration of HASL-based measurements within the class of \emph{Bayesian inference} methods known as  Approximate Bayesian Computation (ABC)~\cite{Marin2011,Sisson2018}. 
Generally speaking  \emph{statistical inference} is interested with inferring  properties of an underlying distribution of probability (in our case $f_\theta$) based on some  data observed through an experiment $y_{exp}$. Bayesian inference methods, on the other hand, rely on  the Bayesian interpretation of probability, therefore starting from some \emph{prior distribution} $\pi(.)$, which expresses an  \emph{initial belief} on the distribution to be estimated over the parameters domain $\Theta$, they allow for computing the \emph{posterior distribution}  $\pi(\theta|y_{exp})$, that is, the target probability distribution over $\Theta$ based on the observed data $y_{exp}$. Formally, in Bayesian statistics,  the posterior distribution is defined by: 
\begin{equation*}
    \pi(\theta|y_{exp}) = \frac{p(y_{exp}|\theta) \pi(\theta)}{\int_{\theta'} p(y_{exp}|\theta') \pi(\theta')\, d\theta'}
\end{equation*}
where $p(.|\theta)$ denotes the \emph{ likelihood function}, that is, the function that measures how probable $y$ is to be observed given  the model's parameters  $\theta$.
An inherent drawback of Bayesian statistics is in that, by definition, the posterior distribution  relies on the accessibility  to  the likelihood function $p(y_{exp}|\theta)$ which,  particularly for complex models,  may be too expensive to compute or even intractable.  ABC algorithms have been introduced  to tackle this issue, i.e. as a \emph{likelihood-free} alternative to classical Bayesian methods (we refer to~\cite{Marin2011,Sisson2018} for  exhaustive surveys of ABC or rejection-sampling methods). The basic idea behind the ABC method is to obtain an estimate, denoted $\pi_{ABC,\epsilon}$, of the posterior distribution $\pi(\theta|y_{exp})$ through an iterative procedure  through which, at each iteration, we draw a parameter vector  $\theta$ from a prior, i.e. $\theta\!\sim\!\pi(.)$ , we simulate the model, and we keep the parameter vector if the corresponding simulation is close enough to the observations according to a threshold $\epsilon$.
 These selected parameters are samples from $\pi_{ABC,\epsilon}$ and approximates the posterior distribution: the smaller the $\epsilon$, the better the approximation. The chosen value of $\epsilon$ is crucial for the performance of ABC algorithm: a small $\epsilon$ is needed to achieve a good approximation, however this may result in high rejection rate leading to cumbersome computations. To overcome this issue, more elaborate algorithms were proposed, like ABC-SMC algorithms \cite{Beaumont2008,DelMoral2012}. 


\section{ABC-HASL method for  tuning oscillators} 
\label{sec:methods}

We introduce an approach for exploring the parameters space of  stochastic oscillators so that a given oscillatory criteria, e.g.,  the mean duration of the oscillation period,  is met.  The approach is based on  the Automata-ABC procedure described in Section~\ref{sec:automata_abc}. The overall idea is to provide one with the characterisation of  some linear hybrid automaton  capable of assessing oscillation related measures and to plug  in the Automata-ABC scheme so that it can effectively be applied to the   analysis the effect the model's parameters have on the oscillations. 
We start off with an overview of preliminary notions  necessary for understanding the functioning of the automata for oscillation related  measures.

\subsection{Characterising of noisy periodicity}
\begin{figure}
\centering
\begin{tabular}{ccc}
\begin{tabular}{c}
\includegraphics[width=0.4\textwidth]
{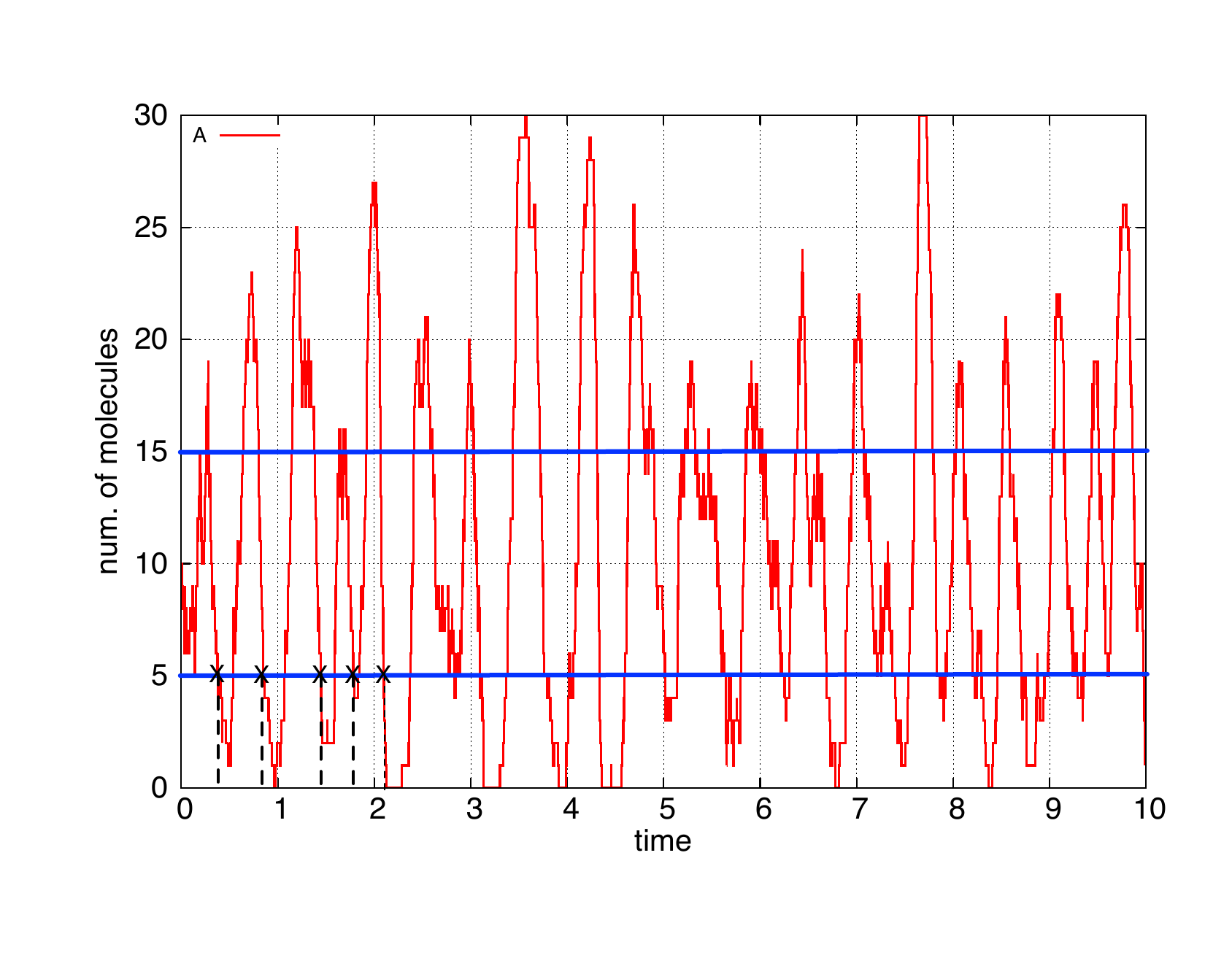} 
\end{tabular} & \hskip 7ex \ &
\begin{tabular}{c}
\includegraphics[width=0.45\textwidth]
{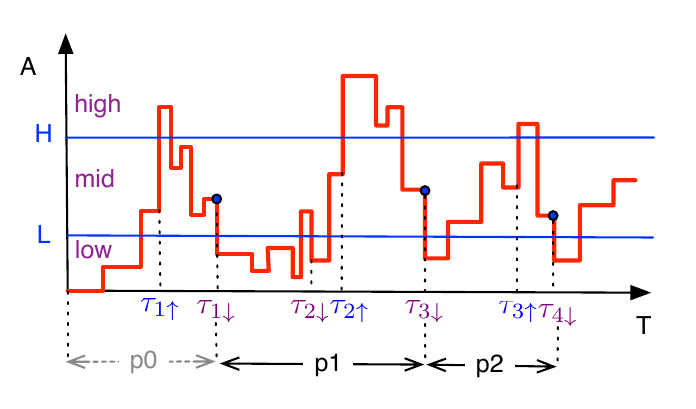}  
\end{tabular}\\
\end{tabular}
   \caption{A \emph{noisy periodic} trajectory (left) and  the corresponding \emph{partition dependent} characterisation of noisy period realisations (right).} 
    \label{fig:noisypertrace}
\end{figure}

The mathematical notion of periodic  function (i.e. a  function $f:\mathbb{R}^+\to\mathbb{R}$ for which  $\exists t_p\in\mathbb{R}^+$ such that $\forall t\in\mathbb{R}^+$, $f(t)=f(t+t_p)$, with $t_p$ being the period) is of little use in the context of  stochastic models as 
paths of a stochastic oscillator   are noisy by nature (e.g. Figure~\ref{fig:noisypertrace}-left) hence will have (unless in degenerative cases) zero probability of matching such strict notion of periodicity.  

Therefore to take into account the  noisy nature of stochastic oscillators we resort to  a less strict   notion of  \emph{noisy periodicity}~\cite{Spieler13}. 
\begin{definition}[noisy periodic trajectory] 
\label{def:noisyperiodicity}
For ${\cal M}_\theta$  an $n$-dimensional DESP  population model 
let  $L,H\IN\mathds{N}$,  $L\!<\! H$, be two levels  establishing  
the  partition $dom_i({\cal M}_\theta)\!=\! low\cup mid\cup high$ with  $low=[0,L)$, $mid=[L,H)$ and $high=[H,\infty)$. 
A path $\sigma\in Path({\cal M}_\theta)$  is said \emph{noisy periodic} w.r.t the $i^{th}$ dimension, and the 
considered  $L,H$ induced partition of $dom_i({\cal M}_\theta)$  if the projection $\sigma_i$ visits the intervals    $low$, $mid$ and $high$ infinitely often.  
\end{definition}
\noindent{\bf H/L-crossing points}. 
Given a  noisy periodic trace $\sigma_A$  we  denote $\tau_{j \downarrow}$ ( $\tau_{j\uparrow}$), 
the instant of time when  $\sigma_A$ enters for the $j$-th time the $low$ ($high$) region. 
 $T_{\downarrow}\!=\!\cup_j \tau_{j\downarrow}$ (resp. $T_{\uparrow}\!=\!\cup_j \tau_{j\uparrow}$) 
is the set of all \emph{low-crossing points} (reps. \emph{high-crossing points}). 
Observe that $T_{\downarrow}$ and $T_{\uparrow}$ 
reciprocally induce a partition on each other. Specifically  $T_{\downarrow}\!=\!\cup_k T_{k\downarrow}$ 
where $T_{k\downarrow}$ is the  subset of  $T_{\downarrow}$ containing 
the $k$-th sequence of contiguous  \emph{low-crossing points} not interleaved by 
any \emph{high-crossing point}. Formally $T_{k\downarrow}\!=\!\{\tau_{i\downarrow}, \ldots, \tau_{(i+h)\downarrow} | \exists k',  \tau_{(i-1)\downarrow}\!<\! \tau_{k'\uparrow}\!<\! \tau_{i\downarrow},  \tau_{(i\!+\!h)\downarrow}\!<\!\tau_{(k'\!+\! 1)\uparrow} \}$. 
Similarly $T_{\uparrow}$ is partitioned  $T_{\uparrow}\!=\!\cup_k T_{k\uparrow}$ where 
$T_{k\uparrow}$ is the  subset of  $T_{\uparrow}$ containing 
the $k$-th sequence of contiguous  \emph{high-crossing points} not interleaved by 
any \emph{low-crossing point}. 
For path $\sigma_A$  in Figure~\ref{fig:noisypertrace} (right) 
we have that $T_{\downarrow}\!=\! T_{1\downarrow}\!\cup\! T_{2\downarrow} \!\cup\! T_{3\downarrow}\ldots $
with $T_{1\downarrow}\!=\! \{\tau_{1\downarrow}, \tau_{2\downarrow}\}$, 
$T_{2\downarrow}\!=\! \{\tau_{3\downarrow}\}$, $T_{3\downarrow}\!=\! \{\tau_{4\downarrow}\}$, 
while $T_{\uparrow}\!=\! T_{1\uparrow}\!\cup\! T_{2\uparrow} \!\cup\! T_{3\uparrow}\ldots $ 
with $T_{1\uparrow}\!=\! \{\tau_{1\uparrow}\}$, $T_{2\uparrow}\!=\! \{\tau_{2\uparrow}\}$, 
$T_{3\uparrow}\!=\! \{\tau_{3\uparrow}\}$. 
Based on   H/L crossing points we formalise the notion of \emph{period realisation} for a noisy period path. 

\begin{definition}[$k^{th}$noisy period realisation]
\label{def:noisyperiod}
For $\sigma_A$  a noisy periodic trajectory with crossing point times $T_{\downarrow}\!=\!\cup_{k\geq 1} T_{k\downarrow}$ , 
respectively $T_{\uparrow}\!=\!\cup_{k\geq 1} T_{k\uparrow}$, 
 the   realisation of the $k^{th}$ noisy period, denoted $t_{p_k}$, is defined as  $t_{p_k}\!=\! min(T_{(k\!+\!1)\downarrow})- min(T_{k\downarrow})$\footnote{$t_{p_k}$ could  alternatively be defined as 
$t_{p_k}\!=\! min(T_{(k\!+\!1)\uparrow})- min(T_{k\uparrow})$, that is,  w.r.t.   crossing into the $high$ region, rather than into 
the $low$ region. 
It is straightforward to show that both definitions are semantically 
equivalent, i.e., the average value of  $t_{p_k}$ measured along a trace  is equivalent with both definitions. }.
\end{definition}
Figure~\ref{fig:noisypertrace} (right) shows  an example of \emph{period realisations}: the  first two period realisations, denoted $p1$ and $p2$, are delimited by the $mid$-to-$low$ crossing points corresponding to the first entering of the $low$ region which follows a previous sojourn in the $high$ region 
and their duration (as per Definition~\ref{def:noisyperiod}) is  $t_{p_1}\!=\! \tau_{3\downarrow}-\tau_{1\downarrow}$
respectively $t_{p_2}\!=\! \tau_{4\downarrow}-\tau_{3\downarrow}$. 
Notice that the time interval denoted as $p0$ does not represent a complete period realisation as  there's no guarantee that   $t=0$ corresponds with the actual entering into the $low$ region.  
Definition~\ref{def:noisyperiod} correctly does not account for the first \emph{spurious} period $p0$. 
Relying on  the notion of   period realisation we characterise  the  \emph{period average} and \emph{period variance} of a noisy periodic trace. 
Observe that the \emph{period variance}
allows us to analyse  the regularity of the observed oscillator, that is, 
a ``regular'' (`irregular'') oscillator is one whose traces exhibit  little (large)  period variance. 
\begin{definition} [period average]
\label{def:averagenoisyperiod}
For $\sigma_A$  a noisy periodic trajectory 
the  period average of the first $n\!\in\! \mathbb{N}$ period realisations, denoted $\overline{t}_{p}(n)$, is defined as $\overline{t}_{p}(n)\!=\!\frac{1}{n}\sum_{k=1}^n t_{p_k}$, 
where $t_{p_k}$ is the $k$-th period realisation. 
\end{definition}

Observe that for a sustained oscillator, the   average value of the noisy-period, in the long run, corresponds to the limit $\overline{t}_{p}=\lim_{n \to \infty}\overline{t}_{p}(n)$. 

\begin{definition}[period variance]
\label{def:flcutnoisyperiod}
For $\sigma_A$  a noisy periodic trajectory 
the  period variance of the first $n\!\in\! \mathbb{N}$ period realisations, denoted $s^2_{t_p}(n)$, is defined as $s^2_{t_p}(n)\!=\!\frac{1}{n-1}\sum_{k=1}^n (t_{p_k}-\overline{t}_{p}(n))^2$, 
where $t_{p_k}$ is the $k$-th period realisation and $\overline{t}_{p}(n)$ is the period average for the first $n$ period realisations. 
\end{definition}
Based on the period average and variance we now introduce a notion of distance of noisy periodic path from a target mean period value. We will employ such  distance in the HASL-based adaptation of the ABC method for inferring the parameters of an oscillator. 
\begin{definition}[distance from target period]
\label{def:distancetargetperiod}
For $\sigma_A$  a noisy periodic trajectory and $\obsmeantp\in\mathbb{R}_{>0}$ a target mean period duration we define the distance of $\sigma_A$ from $\obsmeantp$ w.r.t. the first $n\!\in\! \mathbb{N}$ period realisations as
\begin{equation}
    \label{eq:dist_tp}
    \text{dist}(\sigma_A,n,\obsmeantp)= \text{dist}(\meantp(n), \vartp(n), \obsmeantp) = \min(\frac{\mid\meantp(n)-\obsmeantp\mid}{\obsmeantp}, 
                    \frac{\sqrt{\vartp(n)}}{\obsmeantp})
\end{equation}
where $\meantp(n)$ ($\vartp(n)$) denotes the mean value (the variance) of the first $n$ periods detected along $\sigma_A$ (as per Def.~\ref{def:noisyperiod} and Def.~\ref{def:flcutnoisyperiod}). 
\end{definition}
Notice that distance~(\ref{eq:dist_tp}) establishes a form of multi-criteria selection of parameters as both the mean value and the variance of the detected periods are constrained. For example with a 10\% tolerance (i.e. $\eps=0.1$ in ABC terms)  only the parameters $\theta\in\Theta$ that issue a relative error (w.r.t. the target mean period) not above 0.1 are selected.


\subsection{An automaton for  the distance from a target period}
\begin{figure*}[ht]
\centering
\fbox{
\includegraphics[scale=.39]{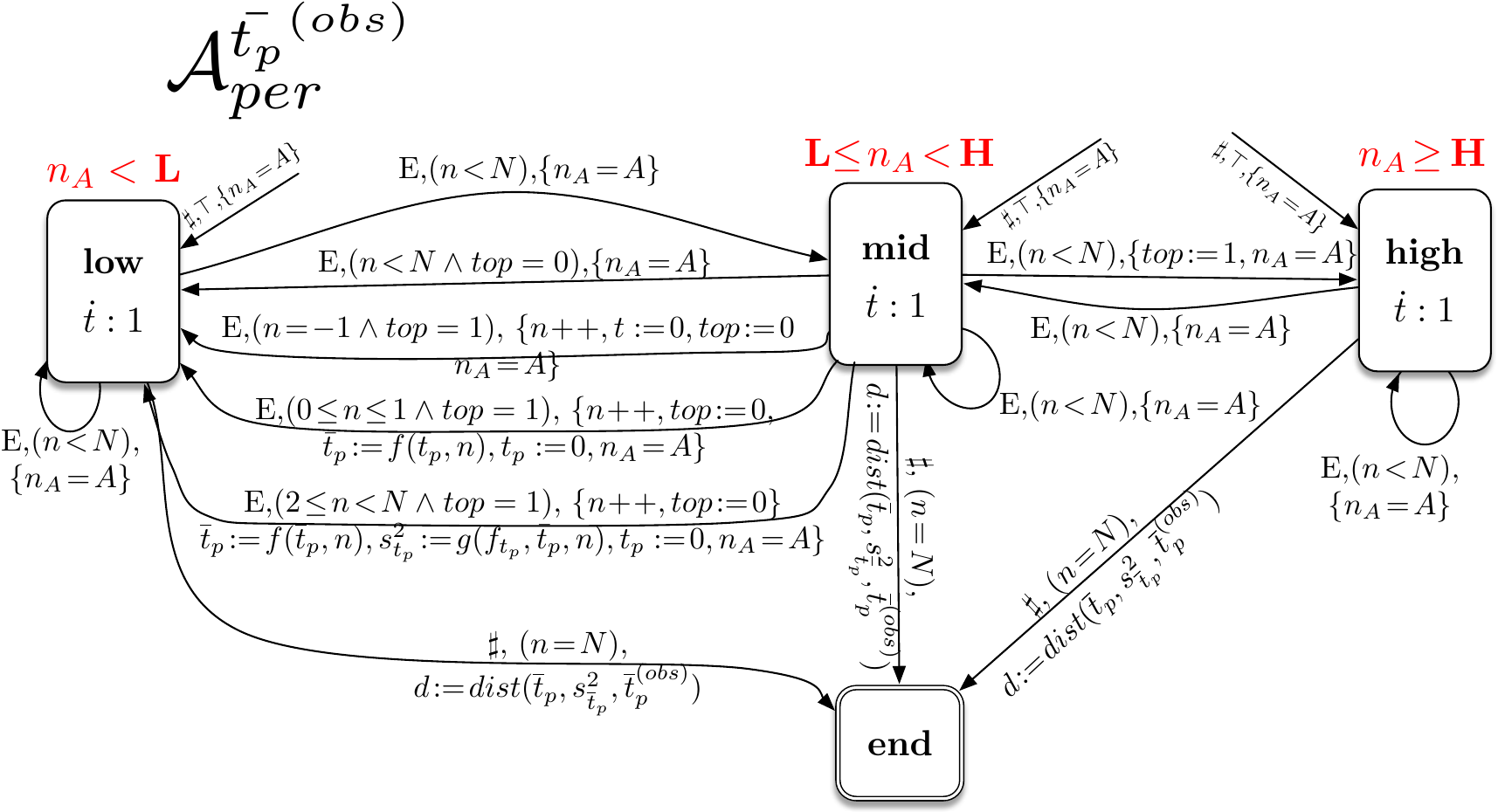}}
\caption{ ${\cal A}_{per}$: an LHA for selecting noisy periodic traces (with respect to an observed species $A$) related to  partition $low=(-\infty,L]$, $mid=(L,H)$ and $high=[H,+\infty)$. }
\label{fig:lhaperiod}
\end{figure*}

We introduce a LHA  named $\autpertpobs$ (Figure~\ref{fig:lhaperiod})    for assessing the distance (as per Definition~\ref{def:distancetargetperiod}) between the mean period measured on paths of DESP oscillator ${\cal M}_\theta$ and a target period duration  $\obsmeantp$. 
The automaton  consists of 
three main locations {\bf low}, {\bf mid} and {\bf high} 
(corresponding to the  regions of the   partition of $A$'s domain induced by thresholds $L<H$). 
Its functioning  is as follows: 
 processing  starts in either of the 3  state (which are all initial) depending on the initial state of the oscillator (population of oscillating species $A$ is stored in $n_A$ which is initialised through autonomous transitions before unfolding of path $\sigma$ begins).  Detection of one period realisation (as of Definition~\ref{def:noisyperiod})  correspond with  the completion of a loop from {\bf low} to  {\bf high} and back to {\bf low}  locations. 
The analysis of the 
simulated trajectory ends by entering location {\bf end} as soon as the $N$-th period has been detected. 
Table~\ref{tab:lhavars2} report about  some of the variables (see~\cite{BentriouPhD} for the complete list) that $\autpertpobs$ uses to store relevant statistics of the simulated path. Variable $d$ is updated with the computed distance (as per Definition~\ref{eq:dist_tp})  on detection of the $N$-th period.  
\begin{table}[]
    \centering\footnotesize
    \begin{tabular}{|c|c|c|c|}
           \hline
    {\bf name} & {\bf domain}  & {\bf update definition} & {\bf description}  \\ \hline
    $n$ & $\mathbb{N}$ & \emph{increment} & counter of  detected periods \\ \hline
    ${t}_p$ & $\mathbb{R}_{\geq 0}$ &  \emph{reset}  & duration  last  period  \\ \hline    
        $\bar{t}_p$ & $\mathbb{R}_{\geq 0}$ & $  f(\bar{t}_p,t_p,n)=\frac{\bar{t}_{p_n}\cdot n+t_p}{n+1}$ & period mean \\ \hline
   $s^2_{t_p}$ & $\mathbb{R}_{\geq 0}$ & $g(s^2_{t_{p}},\bar{t}_p,t_p,n)  =
    \frac{n-1}{n-2}\cdot s^2_{t_{p}}+\frac{(\bar{t}_p-t_p)^2}{n-1}$    & period variance      \\          \hline         
    $d_p$ & $\mathbb{R}_{\geq 0}$ & 
    $\min(\frac{\meantp-\obsmeantp}{\obsmeantp},\frac{\sqrt{\vartp}}{\obsmeantp})$ & distance from target period      \\         
    \hline 
    \end{tabular}
    \caption{ Variables  of the  $\autpertpobs$ automaton. }
    \label{tab:lhavars2}
    \end{table}
\subsection{HASL-ABC method for tuning oscillators}
\label{sec:automata_abc}
To calibrate  the period of noisy oscillators we adapt the HASL-based version of the ABC method~\cite{DBLP:journals/tcs/BentriouBC21} so that it operates with $\aut_{per}$ automaton.\\
\begin{algorithm}[H]
    \begin{algorithmic}
        \REQUIRE $\mcpCTMC$ parametric DESP, $\pi$ prior,
        \\$N$: number of particles, $\eps$: tolerance level, $\autper$ distance period LHA
        \\
        \ENSURE $(\theta^{(i)})_{1\leq i \leq N}$ drawn from $\piabceps$
        \FOR {$i = 1:N$}
        \REPEAT
        \STATE $\theta' \sim \pi$
        \STATE $d' \sim (d_p,\autper) \times \mcCTMC[\theta']$
        \UNTIL {$d' \leq \eps$}
        \STATE $\theta^{(i)} \gets \theta'$
        \ENDFOR
    \end{algorithmic} 
    \caption{Rejection sampling HASL-ABC Algorithm} 
    \label{alg:general_automaton_abc}
\end{algorithm}
\noindent Algorithm~\ref{alg:general_automaton_abc} describes the general HASL-based version of the ABC method. Further from the usual arguments of the ABC scheme (i.e.  a parametric model ${\cal M}_\theta$, a prior distribution over its parameters, the target number of particles $N$ and the level of tolerance $\eps$) it also takes as input an HASL automaton $\aut$ and an expression $Y$ representing a distance measure computed over the paths of the product process ${\cal M}_\theta\times\aut_{per}$. The output is the $N$ values $\theta^{(i)}$ of accepted parameters which represent samples from the posterior distribution. 
By using automaton $\autpertpobs$ and the distance from the target period $\obsmeantp$ as expression $Y\equiv last(d)$ (with $d$  as in Table~\ref{tab:lhavars2}) the algorithm computes an estimation of the posterior w.r.t. to target period. In order to improve the convergence we also developed a sequential version of the HASL-based ABC (see~\cite{Bentriou2008}).  

\section{Case studies}
\label{sec:experiments}
We demonstrate the HASL-ABC procedure for tuning stochastic oscillator on two examples of stochastic oscillators. 

\begin{figure}

\begin{tabular}{cc}
\begin{tabular}{c}
\includegraphics[width=.5\textwidth]{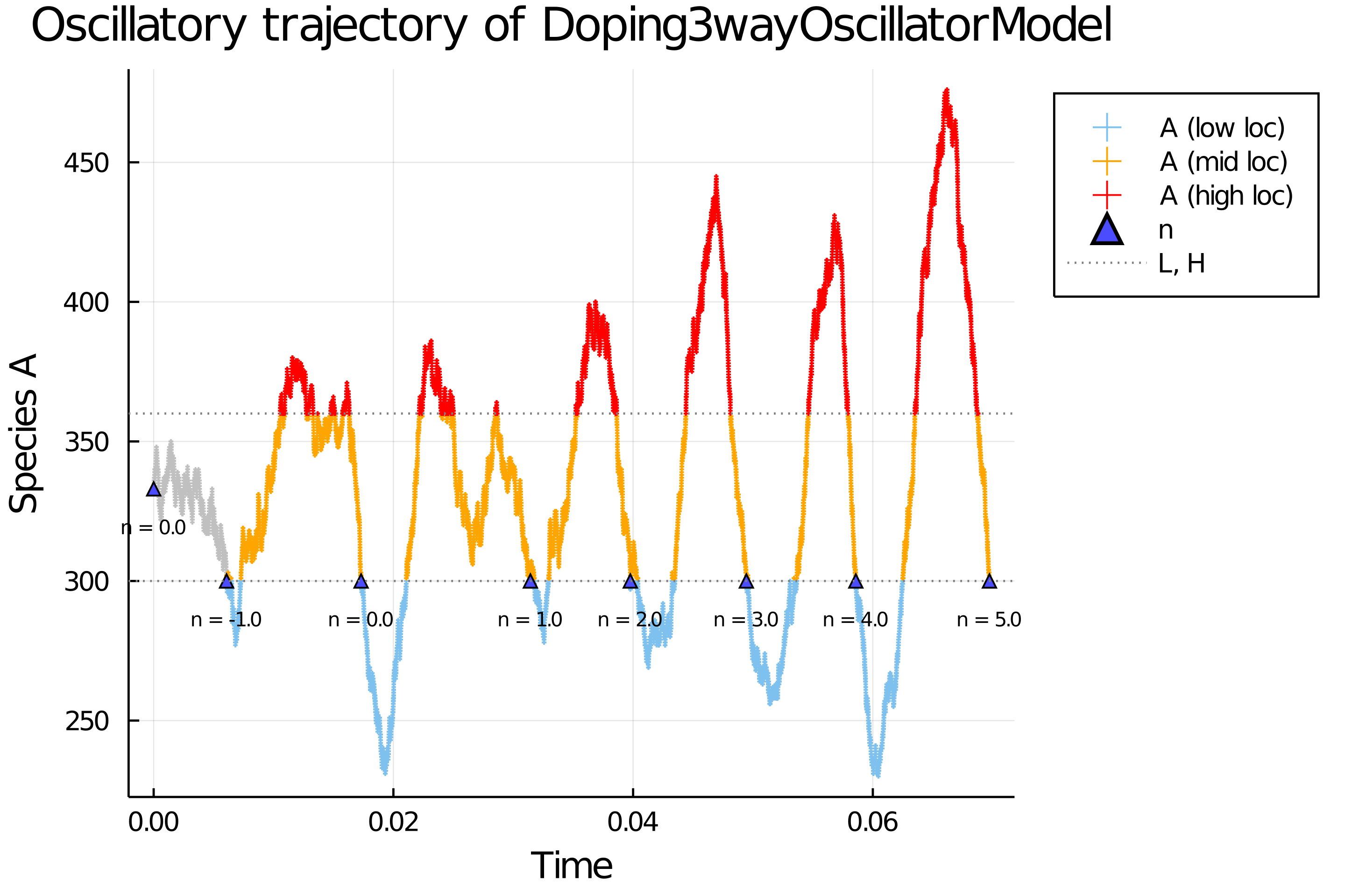}
\end{tabular}
& 
\begin{tabular}{c}
\includegraphics[width=.5\textwidth]{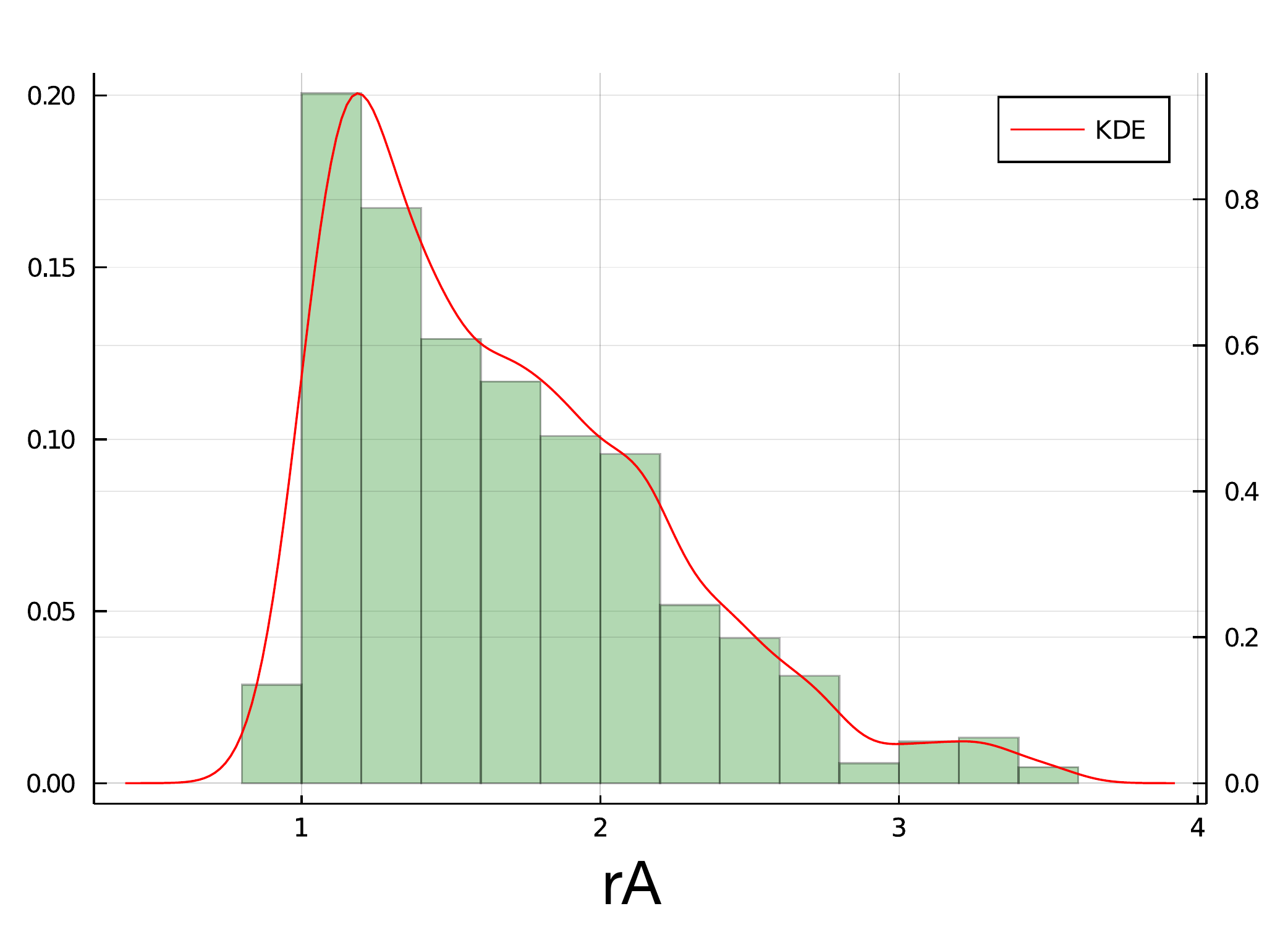}     
\end{tabular}
\end{tabular}
        \caption{The noisy periodic character of species $A$ (left) and the posterior distribution for a single-parameter estimation experiment of the 3-ways oscillator.}
        \label{fig:three graphs}
\end{figure}

\smallskip
\noindent
{\bf A synthetic 3-ways oscillator}.
The CRN in~(\ref{eq:3way_oscillator})  
represents a  model of synthetic sustained oscillator called {\em doping 3-way oscillator}~\cite{Cardelli2009}. 
\begin{gather}
\begin{aligned}
    &R_1\!:\!A \!+\! B \xrightarrow{r_A} 2 B 
    &&R_2\!:\! B \!+\! C \xrightarrow{r_B} 2C 
    && R_3\!:\!C \!+\! A \xrightarrow{r_C} 2A   \\
    &R_4\!:\!D_A \!+\! C \xrightarrow{r_C} D_A\!+\!A 
    &&R_5\!:\! D_B \!+\! A \xrightarrow{r_A} D_B \!+\! B 
    && R_6\!:\!D_C \!+\! B \xrightarrow{r_B} D_C \!+\! C \end{aligned}
    \label{eq:3way_oscillator}
\end{gather}
It consists of 3 main species $A$, $B$ and $C$ forming a positive feedback loop (through reactions $R_1$, $R_2$, $R_3$) plus 3 corresponding invariant (\emph{doping}) species $D_A$, $D_B$, $D_C$ whose goal is to avoid extinction of the main species (through reactions $R_4$, $R_5$, $R_6$). 
It can be shown that the total population  is invariant and that the model yields sustained noisy oscillation for the 3 main species, whose period and amplitude depend on the model's parameters $r_A$, $r_B$ and $r_C$ (as well as on the total population). 
Figure~\ref{fig:three graphs} (left) depicts a simulated trajectory showing the oscillatory character of species A  together with the period realisations induced by a given partition  ($L=300$ and $H=360$). The corresponding  location changes of the synchronised automaton ${\cal A}_{per}$ as the trajectory is unfolded are depicted in different colors.

\begin{figure}
\centering
\begin{tabular}{c|c}
\includegraphics[width=0.5\textwidth]
{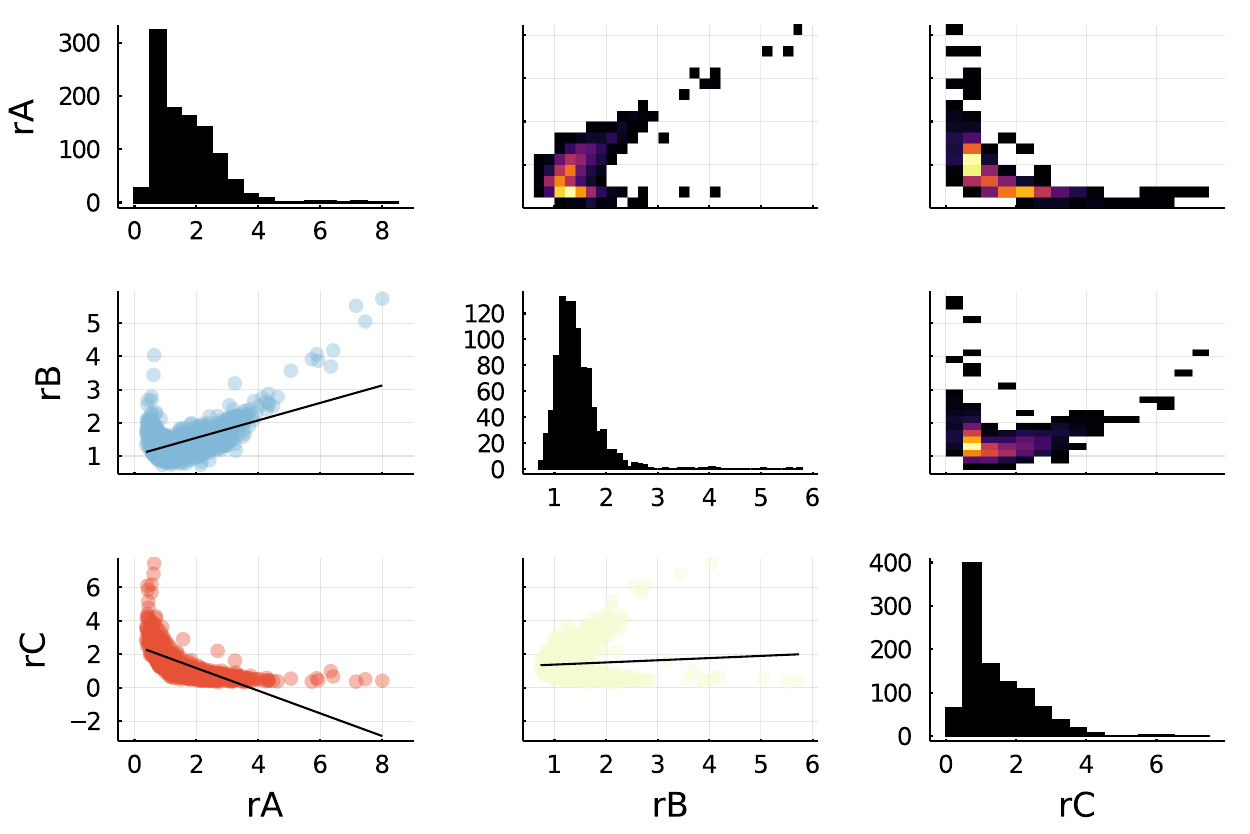}    & \includegraphics[width=0.5\textwidth]
{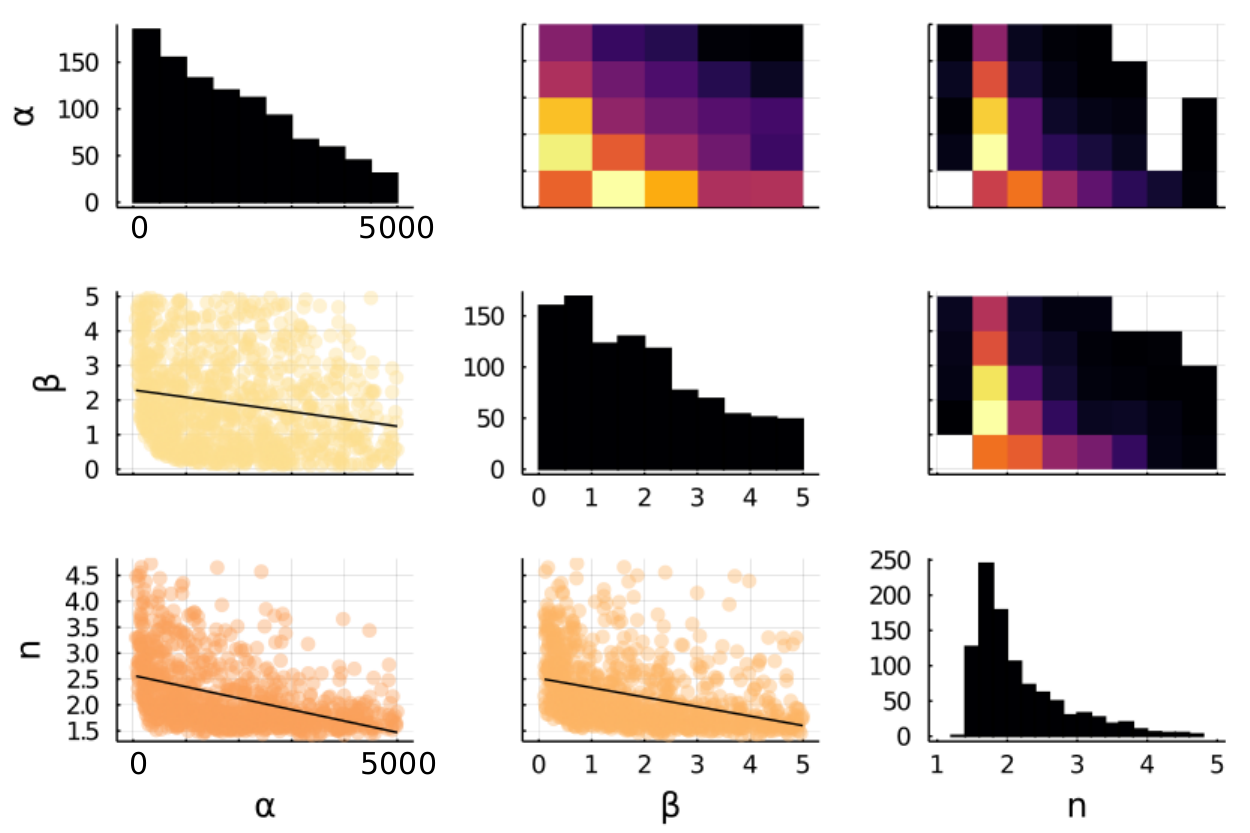}   \\
3-way oscillator & repressilator\\
\end{tabular}
    \caption{Correlation plots of posterior distribution obtained through  $\aut{per}$-ABC  for the 3D experiment of  3-way oscillator (left) and repressilator (right). }
    \label{fig:doping_3d_abc}
\end{figure}

\smallskip
\noindent{\bf Experiment 1}. 
This is a 1-dimensional  experiment in which we considered $\eltS_0 = (A_0, B_0, C_0, (D_A)_0, (D_B)_0, (D_C)_0) = (333, 333, 333, 10, 10, 10)$ as initial state, we  fixed the rate constants $r_B = r_C = 1.0$, and estimated the posterior distribution for $r_A$ considering a uniform $\Unif{0}{10}$ prior and a target mean period  $\obsmeantp = 0.01$.  For the automaton ${\cal A}_{per}$ the noisy-period dependent partition we considered is $L = 300$ and $H = 360$ while for each trajectory we observed $N=4$ periods. For the ABC algorithm we used $N = 1000$ particles and considered a 20\% tolerance ($\epsilon = 0.2$).
Figure~\ref{fig:three graphs} (right) shows the resulting automaton-ABC posterior (histogram and Kernel density estimation) . 
We observe that 1) the posterior support being included in the prior's $[0.0, 4.0]\subset[0,10.0]$, we have reduced the parameter space to a subset where it is probable to obtain trajectories with a mean period of 0.01 (relative to a 20\% tolerance) and 2) the posterior has only one mode, which is quite sensible,  as having fixed $r_B$ and $r_C$, one would expected the mean period duration being directly linked to the kinetics of reaction $R_1$, which is only parametrised by $r_A$.
\noindent{\bf Experiment 2}. 
This is a 3-dimensional version of the previous  experiment in which we considered the same uniform prior for the 3 parameters $r_A,r_b,r_C \sim \Unif{0}{10}$. 
Figure~\ref{fig:doping_3d_abc} (left) shows the correlation plot matrix of the resulting automaton-ABC posterior,  
where each one-dimensional histogram on the diagonal of plot matrix 
represents the marginal distribution of each parameter, whereas  the two-dimensional marginal distribution are given in the upper triangular part of the matrix (e.g.,   plot on position (1,3) refers to  the marginal distribution $p(r_A, r_C | \obsmeantp)$) and  the scatter plots of the two-dimensional marginal distributions are in the  lower triangular matrix. Based on the diagonal plots we observe that most of the parameters that results in a period close to the target one ($\obsmeantp=20$)
are within the support $[0,4] \times [0,3] \times [0,4]$ and, furthermore the particles form a 3D parabolic shape 
since the three two-dimensional projections have a parabolic shape, according to plots in the upper triangular part of Figure~\ref{fig:doping_3d_abc}.
Also, one can notice that for each two-dimensional histogram, the area near the point $(1,1)$ is a high probability area, which is consistent with the previous one-dimensional experiment.

\noindent
{\bf Repressilator}.
We consider an infinite-state model~(\ref{eq:circclock}) of a synthetic genetic network known as Repressilator  developed to reproduce oscillatory behaviours within a cell~\cite{Elowitz2000}.
It consists of 3 proteins $P_1,P_2,P_3$ forming a negative feedback loop, with  $P_1$ repressing $P_2$'s transcription  gene $G_2$,  $P_2$ repressing $P_3$'s  transcription  and so on. 
\begin{small}
\begin{equation}
\label{eq:circclock}
\begin{array}{lclcl}
R1: G_1 \xrightarrow{r_1} G_1 + M_1 & \ \ \ \ 
     & R2: G_2 \xrightarrow{r_2} G_2 + M_2  & \ \ \ \
     &R3: G_3 \xrightarrow{r_3} G_3 + M_3 \\
     R4:  M_1 \xrightarrow{\beta} M_1 + P_1 & \ \ \ \ 
    &R5: M_2 \xrightarrow{\beta} M_2 + P_2  & \ \  \ \ 
    & R6:  M_3 \xrightarrow{\beta} M_3 + P_3  \\
    R7:  M_1 \xrightarrow{1}\emptyset & \ \ \ \ 
    &R8: M_2 \xrightarrow{1} \emptyset & \ \  \ \ 
    & R9:  M_3 \xrightarrow{1} \emptyset  \\
    R10:  P_1 \xrightarrow{1}\emptyset & \ \ \ \ 
    &R11: P_2 \xrightarrow{1} \emptyset & \ \  \ \ 
    & R12:  P_3 \xrightarrow{1} \emptyset  \\    
\end{array}
\end{equation}
\end{small}    

Following~\cite{Elowitz2000} we assumed mass-action dynamics with a common rate constant $\beta$ for translation reactions ($R_4,R_5,R_6$) and  $1$ for species degradation ($R_7$ to $R_{12}$). Conversely transcription reactions ($R1,R_2,R_3$) are assumed to follow a Hill function dynamics given by the follow parameter definitions $r_1 = \frac{\alpha}{1+[P_3]^n} + \alpha_0$, $r_2 = \frac{\alpha}{1+[P_1]^n} + \alpha_0$ and $r_3 = \frac{\alpha}{1+[P_2]^n} + \alpha_0$, where $n$ is the Hill coefficient, $\alpha$ is related to transcription growth and $\alpha_0$ is the parameter related to the minimum level of transcription growth. 
The parameter space is 4-dimensional with $\theta=(\alpha,\beta,n,\alpha_0)\in\mathbb{R}^4$.
Each parameter affects the resulting oscillation 
\noindent{\bf Experiment 1}. 
This is a 3-dimensional  experiment in which we considered $\eltS_0 = ((M_1)_0, (M_2)_0, (M_3)_0, (P_1)_0, (P_2)_0, (P_3)_0) = (0, 0, 0, 5, 0, 15)$ as initial state, we  fixed $\alpha_0=0$, and estimated the posterior distribution for the remaining parameters considering the following priors:  $\alpha \sim \Unif{50}{5000}$, $\beta \sim \Unif{0.1}{5.0}$, $n \sim \Unif{0.5}{5.0}$.
The target mean period was $\obsmeantp = 20$, while the noisy-period dependent partition was  set at $L = 50$ and $H = 200$. 
 For the ABC algorithm we used $N = 1000$ particles and considered a 10\% tolerance ($\epsilon = 0.1$).
Figure~\ref{fig:doping_3d_abc} (right) shows the correlation plot of the resulting automaton-ABC posterior.  
We observe that with this setting the Repressilator oscillations  are most sensitive to parameter $n$ as its   marginal posterior is much narrower than that of $\alpha$ and $\beta$ (i.e. varying $n$ induces more instability than $\alpha$ and $\beta$). More experiments, including a 4 dimensional one, are illustrated in~\cite{BentriouPhD}

\section{Conclusion}
\label{sec:conclusion}
We introduced a methodology that given a parametric, discrete-state stochastic oscillator model, allows for inferring regions of the parameter space that exhibit a positive probability to match a desired mean oscillation period. Such framework relies on a formal characterisation of noisy periodicity 
which is assessed through a meter encoded by a hybrid automaton. Parameter inference is then obtained by plugging of such an automaton-meter within a ABC scheme. 
The added value of such a rather cumbersome  formalism  is in terms of automation, generality and separation of concerns. 
The period meter automaton, being completely  configurable,  can straightforwardly be generated automatically,  therefore  avoiding an annoying overhead to the end user. 
This combined with the fact that the framework inherently takes care of synchronising the model with the automaton results in a highly configurable and generic approach, one in which different oscillation tuning criteria  can easily be taken into account as long as they can be encoded into a corresponding meter automaton. 
Notice that alternative approaches that are not based on formal methods,  such as e.g., those based on auto-correlation analysis, although  effective, are not easily adaptable as they require the implementation of a customised, hard-coded  procedure where periodicity indicators are obtained by offline analysis  of trajectories sampled from the model.

Future developments include the integration of  oscillation-amplitude amongst the criteria for tuning oscillators through a \emph{peak detector} automaton\cite{ballsttt15}. 



\bibliographystyle{plain} 
\bibliography{biblio}

\begin{thebibliography}{10}

\bibitem{DBLP:conf/rocks/AndreychenkoKS12}
Alexandr Andreychenko, Thilo Kr{\"{u}}ger, and David Spieler.
\newblock Analyzing oscillatory behavior with formal methods.
\newblock In Anne Remke and Mari{\"{e}}lle Stoelinga, editors, {\em Stochastic
  Model Checking. Rigorous Dependability Analysis Using Model Checking
  Techniques for Stochastic Systems - International Autumn School, {ROCKS}
  2012, Vahrn, Italy, October 22-26, 2012, Advanced Lectures}, volume 8453 of
  {\em Lecture Notes in Computer Science}, pages 1--25. Springer, 2012.

\bibitem{Aziz1996CSL}
Adnan Aziz, Kumud Sanwal, Vigyan Singhal, and Robert Brayton.
\newblock {Verifying continuous time Markov chains}.
\newblock In Rajeev Alur and Thomas~A. Henzinger, editors, {\em Computer Aided
  Verification}, pages 269--276, Berlin, Heidelberg, 1996. Springer Berlin
  Heidelberg.

\bibitem{Baier2003}
Christel Baier, Boudewijn Haverkort, Holger Hermanns, and Joost-Pieter Katoen.
\newblock {Model-Checking Algorithms for Continuous-Time Markov Chains.}
\newblock {\em Software Engineering, IEEE Transactions on}, 29:524-- 541, 07
  2003.

\bibitem{BALLARINI201553}
P.~Ballarini, B.~Barbot, M.~Duflot, S.~Haddad, and N.~Pekergin.
\newblock Hasl: A new approach for performance evaluation and model checking
  from concepts to experimentation.
\newblock {\em Performance Evaluation}, 90:53 -- 77, 2015.

\bibitem{Ballarini20102019}
P.~Ballarini and M.L. Guerriero.
\newblock Query-based verification of qualitative trends and oscillations in
  biochemical systems.
\newblock {\em Theoretical Computer Science}, 411(20):2019 -- 2036, 2010.

\bibitem{ballsttt15}
Paolo Ballarini.
\newblock Analysing oscillatory trends of discrete-state stochastic processes
  through hasl statistical model checking.
\newblock {\em International Journal on Software Tools for Technology
  Transfer}, pages 1--22, 2015.

\bibitem{DBLP:journals/tcs/BallariniD15}
Paolo Ballarini and Marie Duflot.
\newblock Applications of an expressive statistical model checking approach to
  the analysis of genetic circuits.
\newblock {\em Theor. Comput. Sci.}, 599:4--33, 2015.

\bibitem{Ballarini2009AnalysingChecking}
Paolo Ballarini, Radu Mardare, and Ivan Mura.
\newblock {Analysing Biochemical Oscillation through Probabilistic Model
  Checking}.
\newblock {\em Electronic Notes in Theoretical Computer Science}, 2009.

\bibitem{Beaumont2008}
Mark~A. Beaumont, Jean-Marie Cornuet, Jean-Michel Marin, and Christian~P.
  Robert.
\newblock Adaptive approximate bayesian computation.
\newblock {\em Biometrika}, 96(4):983--990, 2009.

\bibitem{BentriouPhD}
Mahmoud Bentriou.
\newblock {\em Statistical inference and verification of Chemical Reaction
  Networks}.
\newblock PhD thesis, {\'E}cole doctorale Interfaces, University Paris Saclay,
  2021.

\bibitem{bentriou2019}
Mahmoud Bentriou, Paolo Ballarini, and Paul-Henry Courn{\`e}de.
\newblock Reachability design through approximate bayesian computation.
\newblock In {\em International Conference on Computational Methods in Systems
  Biology}, pages 207--223. Springer, 2019.

\bibitem{DBLP:journals/tcs/BentriouBC21}
Mahmoud Bentriou, Paolo Ballarini, and Paul{-}Henry Courn{\`{e}}de.
\newblock Automaton-abc: {A} statistical method to estimate the probability of
  spatio-temporal properties for parametric markov population models.
\newblock {\em Theor. Comput. Sci.}, 893:191--219, 2021.

\bibitem{Bentriou2008}
Mahmoud Bentriou, Stefanella Boatto, Gautier Viaud, Catherine Bonnet, and
  Paul-Henry Courn{\`{e}}de.
\newblock {Assimilation de donn{\'{e}}es par filtrage particulaire
  r{\'{e}}gularis{\'{e}} dans un mod{\`{e}}le d'{\'{e}}pid{\'{e}}miologie.}
\newblock pages 1--6, 2008.

\bibitem{BortolussiMS16}
Luca Bortolussi, Dimitrios Milios, and Guido Sanguinetti.
\newblock {Smoothed model checking for uncertain Continuous-Time Markov
  Chains}.
\newblock {\em Inf. Comput.}, 247:235--253, 2016.

\bibitem{Brim2013ExploringParameterSpace}
Lubo{\v{s}} Brim, Milan {\v{C}}e{\v{s}}ka, Sven Dra{\v{z}}an, and David
  {\v{S}}afr{\'a}nek.
\newblock Exploring parameter space of stochastic biochemical systems using
  quantitative model checking.
\newblock In Natasha Sharygina and Helmut Veith, editors, {\em Computer Aided
  Verification}, pages 107--123, Berlin, Heidelberg, 2013. Springer Berlin
  Heidelberg.

\bibitem{Cardelli2009}
Luca Cardelli.
\newblock {\em Artificial Biochemistry}, pages 429--462.
\newblock Springer Berlin Heidelberg, Berlin, Heidelberg, 2009.

\bibitem{Ceska2014ParameterSynthesis}
Milan {\v{C}}e{\v{s}}ka, Frits Dannenberg, Marta Kwiatkowska, and Nicola
  Paoletti.
\newblock Precise parameter synthesis for stochastic biochemical systems.
\newblock In Pedro Mendes, Joseph~O. Dada, and Kieran Smallbone, editors, {\em
  Computational Methods in Systems Biology}, pages 86--98, Cham, 2014. Springer
  International Publishing.

\bibitem{ChabrierRivier2004ModelingAQ}
Nathalie Chabrier-Rivier, Marc Chiaverini, Vincent Danos, Fran{\c c}ois Fages,
  and Vincent Sch{\"a}chter.
\newblock Modeling and querying biomolecular interaction networks.
\newblock {\em Theor. Comput. Sci.}, 325:25--44, 2004.

\bibitem{DBLP:conf/popl/ClarkeES83}
Edmund~M. Clarke, E.~Allen Emerson, and A.~Prasad Sistla.
\newblock Automatic verification of finite state concurrent systems using
  temporal logic specifications: {A} practical approach.
\newblock In John~R. Wright, Larry Landweber, Alan~J. Demers, and Tim
  Teitelbaum, editors, {\em Conference Record of the Tenth Annual {ACM}
  Symposium on Principles of Programming Languages, Austin, Texas, USA, January
  1983}, pages 117--126. {ACM} Press, 1983.

\bibitem{DelMoral2012}
Pierre Del~Moral, Arnaud Doucet, and Ajay Jasra.
\newblock {An adaptive sequential Monte Carlo method for approximate Bayesian
  computation}.
\newblock {\em Statistics and Computing}, 22(5):1009--1020, 2012.

\bibitem{Elowitz2000}
M.~Elowitz and S.~Leibler.
\newblock A synthetic oscillatory network of transcriptional regulators.
\newblock {\em Nature}, 403(335), 2000.

\bibitem{Goldbeter2002ComputationalAT}
Albert Goldbeter.
\newblock Computational approaches to cellular rhythms.
\newblock {\em Nature}, 420:238--245, 2002.

\bibitem{Han2008ParameterSynthesis}
T.~{Han}, J.~{Katoen}, and A.~{Mereacre}.
\newblock Approximate parameter synthesis for probabilistic time-bounded
  reachability.
\newblock In {\em 2008 Real-Time Systems Symposium}, pages 173--182, 2008.

\bibitem{MacLaurin22}
Thomas PJ Lindner~B. MacLaurin~J, Fellous~JM.
\newblock Stochastic oscillators in biology: introduction to the special issue.
\newblock {\em Biol Cybern.}, (116(2)):119--120., 2022.

\bibitem{Marin2011}
Jean-Michel Marin, Pierre Pudlo, Christian~P. Robert, and Robin~J. Ryder.
\newblock Approximate bayesian computational methods.
\newblock {\em Statistics and Computing}, 22(6):1167--1180, Nov 2012.

\bibitem{DBLP:conf/cmsb/MolyneuxA20}
Gareth~W. Molyneux and Alessandro Abate.
\newblock Abc(smc)\({}^{\mbox{2}}\): Simultaneous inference and model checking
  of chemical reaction networks.
\newblock In Alessandro Abate, Tatjana Petrov, and Verena Wolf, editors, {\em
  Computational Methods in Systems Biology - 18th International Conference,
  {CMSB} 2020, Konstanz, Germany, September 23-25, 2020, Proceedings}, volume
  12314 of {\em Lecture Notes in Computer Science}, pages 255--279. Springer,
  2020.

\bibitem{molyneux2020bayesian}
Gareth~W. Molyneux, Viraj~B. Wijesuriya, and Alessandro Abate.
\newblock Bayesian verification of chemical reaction networks, 2020.

\bibitem{DBLP:conf/focs/Pnueli77}
Amir Pnueli.
\newblock The temporal logic of programs.
\newblock In {\em 18th Annual Symposium on Foundations of Computer Science,
  Providence, Rhode Island, USA, 31 October - 1 November 1977}, pages 46--57.
  {IEEE} Computer Society, 1977.

\bibitem{Sisson2018}
Scott~A Sisson, Yanan Fan, and Mark Beaumont.
\newblock {\em Handbook of approximate Bayesian computation}.
\newblock Chapman and Hall/CRC, 2018.

\bibitem{doi:10.1073/pnas.0506135103}
J.~Sneyd, K.~Tsaneva-Atanasova, V.~Reznikov, Y.~Bai, M.~J. Sanderson, and D.~I.
  Yule.
\newblock A method for determining the dependence of calcium oscillations on
  inositol trisphosphate oscillations.
\newblock {\em Proceedings of the National Academy of Sciences},
  103(6):1675--1680, 2006.

\bibitem{Spieler13}
D.~Spieler.
\newblock Characterizing oscillatory and noisy periodic behavior in markov
  population models.
\newblock In {\em {{P}roc. {QEST}'13}}, 2013.

\end{thebibliography}


\end{document}